\documentclass[superscriptaddress,twocolumn,prl,showpacs]{revtex4}
\usepackage{amsfonts}
\usepackage{amsmath}
\usepackage{amssymb}
\usepackage{graphicx}

\begin{document}

\preprint{}
\title[Short title for running header]{Storing images in warm atomic vapor}
\author{M. Shuker}
\affiliation{Department of Physics, Technion-Israel Institute of Technology, Haifa 32000,
Israel}
\author{O. Firstenberg}
\affiliation{Department of Physics, Technion-Israel Institute of Technology, Haifa 32000,
Israel}
\author{R. Pugatch}
\affiliation{Department of Physics of Complex Systems, Weizmann Institute of Science,
Rehovot 76100, Israel }
\author{A. Ron}
\affiliation{Department of Physics, Technion-Israel Institute of Technology, Haifa 32000,
Israel}
\author{N. Davidson}
\affiliation{Department of Physics of Complex Systems, Weizmann Institute of Science,
Rehovot 76100, Israel }

\begin{abstract}
Reversible and coherent storage of light in atomic medium is a promising
method with possible application in many fields. In this work, arbitrary
two-dimensional images are slowed and stored in warm atomic vapor for up to
30 $\mu$s, utilizing electromagnetically induced transparency. Both the
intensity and the phase patterns of the optical field are maintained. The
main limitation on the storage resolution and duration is found to be the
diffusion of atoms. A technique analogous to phase-shift lithography is
employed to diminish the effect of diffusion on the visibility of the
reconstructed image.
\end{abstract}

\maketitle

When a resonant light pulse ("probe") impinges upon a gas of atoms, it is
strongly absorbed, exciting the atoms to an upper state. However, if a
second \textquotedblleft pump\textquotedblright\ beam is present, which
couples a second state to the same excited state, than the \textquotedblleft
probe\textquotedblright\ pulse will be able to pass the sample - a
phenomenon known as electromagnetically induced transparency (EIT) \cite%
{HarisEIT1997}. The unique properties of EIT\ allow for a wide range of
coherent light-matter phenomena, including non-linear optics \cite%
{HarisPRL1999_NonLinear}, entanglement \cite%
{LukinPRL2000NonLinearAndEntangelement}, generation of quantum pulses of
light \cite{LukinPRL2004_ShapingQuantumLightPulses}, and quantum
communication \cite{QuantumCommunicationNature2001}. In Ref. \cite%
{HauSlowing1999}, the group velocity of the probe pulse, in a medium of
ultra-cold atoms, was decreased to $17$ m/s and similar results were
achieved in warm vapor \cite{ScullyPRL1999UlraslowGroupVelocity}. In Refs.
\cite{LukinDarkStatePolaritons1,HauStorage2001,LukinPRL2001}, it was
demonstrated that the probe pulse can be completely stopped, while it is
contained inside the medium, by shutting off the pump. The pulse can be
recovered by reopening the pump beam after a certain "storage duration". A
prominent feature of this technique is the reversible and coherent storage
of the information carried by the probe, in the atomic coherences.

Here, we report a method for reversibly capturing complex three-dimensional
light fields using EIT in atomic vapor \cite%
{HarisEIT1997,LukinDarkStatePolaritons1}. The storage experiments discussed
above have utilized a transverse Gaussian mode for both pump and probe
beams. The current research is focused on slowing and storing information
imprinted in the transverse plane of the probe beam ("images"), and on
reducing the effects of atomic diffusion. In a previous work \cite%
{Howell2007_slowing_images}, the ability to slow images and delay them for
several ns was demonstrated, using dispersion from far-detuned absorption
lines. In the current work, we use EIT to slow images to a group velocity of
several thousands m/s, achieving delays of several $\mu$s. We further use
the unique properties of EIT to store the images in the atomic medium for a
similar duration. The long slowing delays and storage durations are
comparable with the typical diffusion time in which atoms cross the image.
We demonstrate the deteriorating effect of diffusion by storing images of
digits for different durations. Finally, we introduce a technique to
diminish the effect of atomic diffusion by alternating the phase of
neighboring features. This technique, which is the atomic analogue of the
optical phase-shift lithography \cite{LevensonPSL}, is demonstrated by
storing an image of three lines and studying the effect of flipping the
phase of adjacent lines.

Slowing of images may prove useful for various image processing and
image-correlation applications. An intriguing possibility is to use the
diffusion during slowing or storage as a complex diffusion filter, thus
realizing an all optical edge detector \cite{GilboaComplexDiffusion}. In the
context of quantum information processing, a single-cell device for
multi-qubit memory can be implemented, e.g., by a spatial array of optical
qubits in the transverse plane. Storing images, rather than slowing them,
has the fundamental advantages of supporting longer delays and completely
converting the optical information to atomic coherence, hence making it more
amenable to various methods of manipulation, e.g., phase conjugation \cite%
{ZibrovPRL2002_Phase_Conjugation}.

\begin{figure}[ptb]
\begin{center}
\includegraphics
{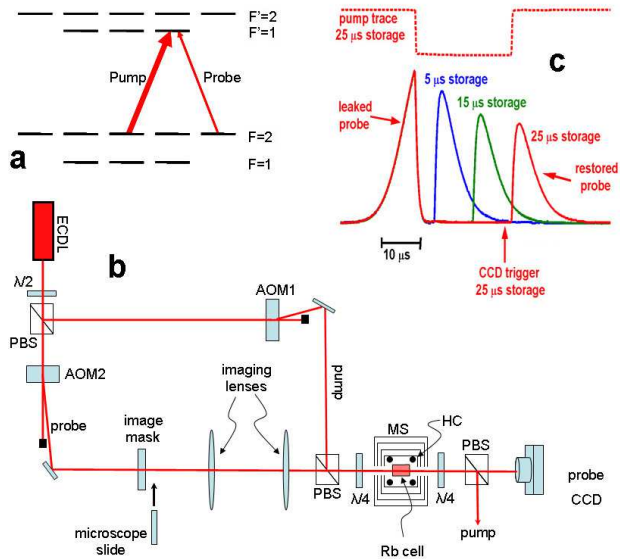}
\end{center}
\caption{The experimental setup and sequence. a) The energy level scheme of
the $^{87}$Rb D1 transition and the transitions of the pump and the probe.
b) The experimental setup: ECDL - external cavity diode laser, $\protect%
\lambda$/2 - half wave-plate, PBS - polarizing beam splitter, AOM - acousto
optic modulator, $\protect\lambda$/4 - quarter wave-plate, MS - magnetic
shield, HC - Helmholtz coils. c) The experimental sequence. The probe is
shaped as a temporal Gaussian with $\protect\sigma$=5$\protect\mu$s. After
about half of the Gaussian pulse leaves the vapor cell ("leaked probe"), the
remaining half is stored in the medium by turning off the pump beam. After a
certain storage duration, the pump is turned back on, and the second half of
the probe leaves the cell ("restored probe"). The figure shows probe
detector traces for storage durations of 5, 15, and 25 $\protect\mu$s and
the pump trace for the 25 $\protect\mu$s case.}
\label{fig_exp_setup}
\end{figure}

The\ storage experiments were performed within the D1 transition of $^{87}$%
Rb \cite{LukinPRL2001} as depicted in Fig. \ref{fig_exp_setup}.a. Two Zeeman
sub-levels of the ground state ($\left\vert F=2;m_{F}=0\right\rangle
,\left\vert F=2;m_{F}=+2\right\rangle $) are used as the two lower levels of
a $\Lambda-$system. The experimental setup is depicted in Fig. \ref%
{fig_exp_setup}.b. An external cavity diode laser (ECDL) is stabilized to
the $F=2\rightarrow F^{\prime}=1$ transition. The laser is divided into two
beams of perpendicular polarizations, the pump and the probe, using a
polarizing beam-splitter (PBS). The pump and the probe pass through
acousto-optic modulators (AOM), allowing us to precisely control the Raman
detuning. The pump beam is shaped as a large Gaussian beam with a waist
radius of $w_{\text{pump}}=2.1$ mm and a total intensity of $5.8$ mW. The
probe beam is shaped as a Gaussian beam with a waist radius of $w_{\text{%
probe}}$=1 mm and intensity of 200 $\mu$W and passes through a binary image
mask, which is a standard resolution target. The plane of the image-mask is
imaged onto the center of the vapor cell using two lenses (the optical
design and the feature size of the images insure that the diffraction along
the cell's length is negligible). The pump and the probe are recombined on a
second PBS and co-propagate towards the cell. A quarter wave-plate before
the cell converts the pump and the probe polarizations to $\sigma^{+}$ and $%
\sigma^{-}$ respectively. We use a $5$ cm long vapor cell, containing
isotopically pure $^{87}$Rb and $10$ Torr of Neon buffer gas. Frequent
collisions with the buffer gas induce Dicke-like narrowing \cite%
{FirstenbergPRA2007_DickeCPTTheory}, increasing the acceptance angle of the
probe beam \cite{WeitzPRA2005Dicke,ShukerPRA2007DickeAngle}, thus allowing
for smaller features in the transverse plane. The diffusion coefficient of
Rb atoms is determined from independent measurements to be $D=10$cm$^{2}$/s.
The temperature of the cell is $\sim52^{\circ}$ C, providing a Rubidium
vapor density of $\sim1.3\times10^{11}$ /cc. The cell is placed within a
four-layered magnetic shield, and a set of Helmholtz coils provides a small,
$B_{z}=50$ mG, axial magnetic field to set the quantization axis.

\begin{figure}[ptb]
\begin{center}
\includegraphics
{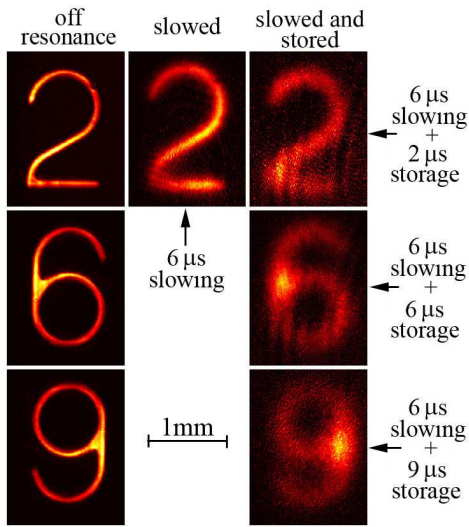}
\end{center}
\caption{Images of digits that were slowed and stored for different
durations in warm vapor. The left column shows the original images that were
to the cell. The middle column shows the image of the digit '2' that was
slowed for about 6$\protect\mu$s. The right column shows the images after
they were stored and retrieved. The images of the digits '2', '6', and '9'
were stored for 2, 6, and 9 $\protect\mu$s, respectively. The decay rate in
the experiments was about $14000$ s$^{-1}$, and the total intensity in all
the experimental data was normalized for the clarity of presentation.}
\label{fig_digits}
\end{figure}

The experimental sequence of the storage experiments is detailed in Fig. \ref%
{fig_exp_setup}.c. A pulse of the probe beam (Gaussian; $\sigma$=$5\mu$s)
travelled in the EIT medium with a group velocity of $\sim8000$ m/s ---
resulting in a \emph{slowly propagating image }\cite%
{Howell2007_slowing_images}. The images were stored in the medium by turning
off the pump after about half of the probe pulse had exit the cell. After an
arbitrary storage duration, the pump beam was turned on, and the stored
image reappeared. Upon exiting the cell, the beams were separated using
polarization optics, and the probe beam was imaged onto a CCD camera, whose
focus was adjusted to the exit plane of the cell. The camera trigger was set
to measure only the restored part of the probe beam (see Fig. \ref%
{fig_exp_setup}.c), and special care is taken to avoid the detection of the
leaked part of the probe.
As a proof-of-principle, we stored images of the digits '2', '6', and '9' in
the atomic vapor. The images were created using a standard resolution target
as a mask, and by shifting the mask to imprint different digits in different
measurements. Fig. \ref{fig_digits} shows the images without slowing (left
column), with slowing only (center column) and after being stored and
retrieved (right column). Evidently, the image of the digit '2', which was
slowed for 6 $\mu$s, was smoothed, and its sharp edges were softened
(similar images were achieved for the other digits). We attribute this
effect to atomic diffusion during the slow propagation of the dark-state
polariton \cite{Pugatch2007_Helical}. The diffusion effect is pronounced due
to the sharp and narrow features in the image and due to the significantly
larger delay time, compared with a previous demonstration of slow images in
hot vapor \cite{Howell2007_slowing_images}. The lower signal-to-noise ratio
in the slowed image is mainly due to the lower transmission, compared with
the off-resonance image.

\begin{figure}[ptb]
\begin{center}
\includegraphics
{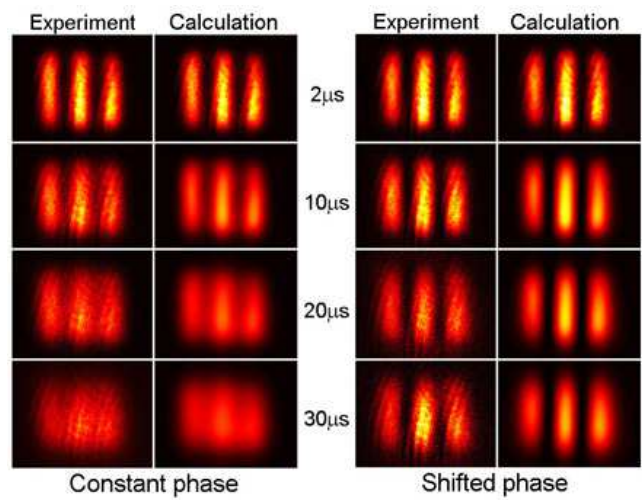}
\end{center}
\caption{Storage of an image of three resolution lines for up to 30 $\protect%
\mu$s. The left hand side shows experimental results (left column) and
theoretical prediction (right column), when the image was sent to the cell
with a constant phase. A good agreement between the experiment and the
calculation was obtained. The right hand side shows the improved resolution
achieved when a $\protect\pi$ phase-shift was applied to the two outer lines
in the image.}
\label{fig_lines_const_pi_exp_cal}
\end{figure}

The images of the digits '2','6', and '9' were stored for 2, 6, and 9 $\mu$%
s, respectively. During the storage, the images are encoded in the quantum
state of the atomic ensemble \cite{LukinDarkStatePolaritons1} and thus
susceptible to degradation through various relaxation processes, such as
spin-exchange, collisions, and diffusion. Nevertheless, the main features
survive through the storage process, and the digits are still identifiable.
Two degradation effects are evident from the stored images: broadening of
the image features due to diffusion and decreased signal-to-noise ratio due
to various decay mechanisms and technical limitations.
In order to study quantitatively the limiting factors of the spatial
resolution, we stored an image of three resolution lines. Both the lines'
thickness and spacing were 340 $\mu$m at the center of the cell. The left
side of Fig. \ref{fig_lines_const_pi_exp_cal} shows the experimental results
for storage durations of $2,~10,~20~$, and $30$ $\mu$s (left column) as well
as the theoretical prediction (right column). For storage time of $2$ $\mu$%
s, the lines are clearly visible, and, as storage time increases to $30$ $\mu
$s, the image blurs and its visibility diminishes. The theoretical
prediction is based on the assumption that, for a plane-wave pump, the
storage of a weak probe field effectively maps its envelope, $E(\mathbf{r},0)
$, onto the spatially-dependent slowly-varying ground-state coherence, $%
\sigma_{bc}\left( \mathbf{r},0\right) $ \cite{LukinDarkStatePolaritons1}.
Assuming that the internal state of each individual atom does not change as
a result of diffusion, we find that the motion of the atoms during storage
can be effectively described by adding a classical diffusion term, $%
D\nabla^2\sigma_{bc}$, to the Bloch equation, where $D$ is the spatial
diffusion coefficient \cite{FirstenbergThermalMotion}. The ensemble average
of the different atoms arriving to the same small macroscopic volume
determines the retrieved probe field \cite{WalsworthRamseyNarrowing2006}.
Therefore, retrieved fields after storage durations $t$ and $t+\tau$ are
related by:
\begin{equation}
\left\vert E\left( \mathbf{r},t+\tau\right) \right\vert ^{2}=\left\vert e^{-%
\frac{\Gamma}{2}\tau}\int d^{3}\mathbf{\tilde{r}}\left\vert E\left( \mathbf{%
\tilde{r}},t\right) \right\vert e^{i\phi\left( \mathbf{\tilde{r}},t\right)
}G_\tau(\mathbf{r-\tilde{r}})\right\vert ^{2},
\end{equation}
where $G_\tau(\mathbf{r})=(2\pi D\tau)^{-3/2}\exp[-\left\vert \mathbf{r}%
\right\vert ^{2}/\left( 4D\tau\right) ]$ is the three-dimensional diffusion
propagator, $\Gamma$ is the ground-state decay rate, $E\left( \mathbf{\tilde{%
r}},t\right) =|E\left( \mathbf{\tilde{r}},t\right) |e^{i\phi\left( \mathbf{%
\tilde{r}},t\right) }$, and $\phi\left( \mathbf{\tilde{r}},t\right) $ is the
phase pattern. Eventually, the initial conditions for the calculation in
Fig. \ref{fig_lines_const_pi_exp_cal} are $\left\vert E\left( \mathbf{\tilde{%
r}},t=2\mu s\right) \right\vert$, obtained by taking the \emph{square-root}
of the measured intensity image after storage of 2 $\mu$s. We used $%
\phi\left( \mathbf{\tilde{r}},t\right)=0 $ for the constant-phase
experiments and an appropriate phase pattern for experiments where the phase
of different areas was shifted (see below).

The diffusion of atoms during the storage is evidently the main limit on the
storage and retrieval resolution in a medium of warm atoms. A possible
method to increase the spatial resolution is based on the fact that the
phase pattern of the light field is stored on top of its intensity
distribution. By appropriately manipulating the phase of different points in
the image, the immunity to diffusion can be improved. In the field of
photolithography, a well established technique to reduce the effect of
spreading due to diffraction is the so-called phase-shift lithography \cite%
{LevensonPSL}. Its concept is based on flipping the phases of neighboring
features of the image, so that light diffracting to the areas between them
will interfere destructively. We implemented an analogous technique in order
to mitigate the effect of atomic diffusion. Using microscope slides, we
shifted the phase of the two outer lines by $\sim\pi$ with respect to the
central line in the probe image. Therefore, the ground-state coherence,
created in the atomic medium, had opposite phases in neighboring lines.
During storage, atoms of opposite phases diffused to the area between the
lines, and the amplitude of the coherence field sum up to zero. Therefore,
no probe beam was generated in the forward direction \cite%
{YelinDecoherenceVortices}, keeping these areas dark in the restored image.

\begin{figure}[ptb]
\begin{center}
\includegraphics
{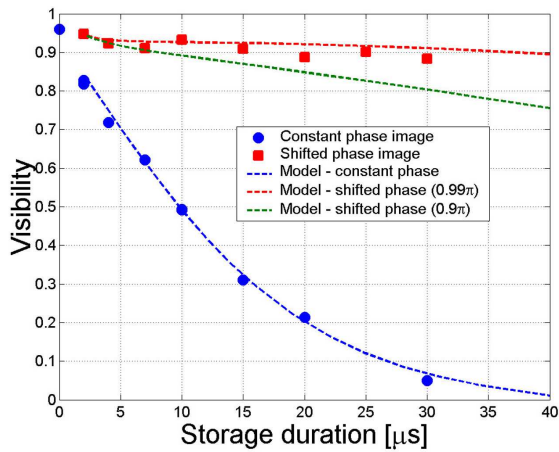}
\end{center}
\caption{The visibility of the lines in Fig. \protect\ref%
{fig_lines_const_pi_exp_cal} as a function of storage duration. The
visibility of the constant phase image drops rapidly due to diffusion, while
the shifted phase image shows only a small decrease in the visibility. The
results are well described by our theoretical model, taking into account
inaccuracies in the phase pattern of the phase-shift experiment. An error of
$0.01\protect\pi$ in the phase of the left-most line gives the best fit to
the experimental data (red dashed curve). A curve for an error of $0.1%
\protect\pi$ is also plotted (green dashed curve) to demonstrate the effect
of phase inaccuracies. The visibility presented here is obtained from the
two left-most lines, because of a relatively large error in the phase of the
right-most line (estimated to be $0.2\protect\pi$).}
\label{fig_vis_pi_const}
\end{figure}

The results of the shifted-phase experiment as well as suitable calculations
are depicted in the right side of Fig. \ref{fig_lines_const_pi_exp_cal}. The
calculation is performed by introducing an appropriate stepwise phase
pattern. The dramatic effect of the phase-shift on the visibility of the
stored features is pronounced --- even after 30 $\mu$s, the lines are well
separated, while the respective images in the constant-phase case are
completely diffused. These results demonstrate that indeed a \emph{complex
field} is stored in the atomic medium. We have previously demonstrated the
topological stability of stored optical vortices to diffusion \cite%
{Pugatch2007_Helical}. The current method is a generalized technique that
can be adapted to \emph{arbitrary} images. Figure \ref{fig_vis_pi_const}
shows the measured and calculated visibility of the stored image as a
function of the storage duration. The rapid decrease of visibility in the
constant phase experiment as well as the persistent high visibility in the
shifted-phase experiment are well described by our theoretical model.

The phase-shift method can be applied in cases where there is
either partial or full a-priori information about the image. In
the demonstration above, a prior knowledge regarding the location
of the lines in the image allowed us to construct a suitable phase
mask. If complete information regarding the image is available, an
elaborate phase mask can be designed, even for complicated images,
to minimize the effects of diffusion \cite{LevensonPSL}. If no
a-priori information regarding the image is available, other
techniques can be used to reduce the effect of diffusion.
Increasing the buffer gas pressure is a straight-forward option,
limited by the one and two-photon broadening mechanisms. In this
case the phase pattern can be used to store information, which
might be suitable for certain quantum memory realizations
\cite{FliLukPRA2002QuantumMemoryForPhotons,LukinRMP2003}. Another
novel method to reduce the effects of diffusion on arbitrary
stored images, based on storing the \emph{transform plane}, was
recently investigated theoretically \cite{Yelin4fStorage}, and
implemented experimentally \cite{HowellPRL2008}.

In conclusion, we have demonstrated the ability to store images in warm
atomic vapor. The main limitation of the spatial resolution is the diffusion
of atoms during the storage. A technique to reduce the effect of diffusion
by alternating the phase in neighboring features of the image has been
demonstrated by storing 340 $\mu$m thick lines for as long as 30 $\mu$s,
with negligible decrease in their visibility. Our experiment is a trivial
case of storing a three-dimensional complex light field, where the shape in
the longitudinal (temporal) direction is a simple Gaussian. In principle, it
should be possible to store more elaborate temporal shapes, thus storing
three dimensional information ("movie"). \newline
Helpful discussions with Paz London and Amit Ben-kish are acknowledged. This
work was partially supported by the fund for encouragement of research in
the Technion.

\bibliographystyle{apsrev}
\bibliography{references_imaging}

\end{document}